\documentclass[useAMS,usenatbib,epsfig]{mn2e}

\pdfoutput=1
\usepackage{natbib}
\usepackage{url}
\usepackage{amsmath,amsfonts}
\usepackage{epsfig}
\usepackage{amssymb}
\usepackage{rotating}

\def\reff@jnl#1{{\rm#1\/}}

\def\aj{\reff@jnl{AJ}}                  
\def\araa{\reff@jnl{ARA\&A}}            
\def\apj{\reff@jnl{ApJ}}                        
\def\apjl{\reff@jnl{ApJ}}               
\def\apjs{\reff@jnl{ApJS}}              
\def\ao{\reff@jnl{Appl.Optics}}         
\def\apss{\reff@jnl{Ap\&SS}}            
\def\aap{\reff@jnl{A\&A}}               
\def\aapr{\reff@jnl{A\&A~Rev.}}         
\def\aaps{\reff@jnl{A\&AS}}             
\def\azh{\reff@jnl{AZh}}                        
\def\baas{\reff@jnl{BAAS}}              
\def\jrasc{\reff@jnl{JRASC}}            
\def\memras{\reff@jnl{MmRAS}}           
\def\mnras{\reff@jnl{MNRAS}}            
\def\pra{\reff@jnl{Phys.Rev.A}}         
\def\prb{\reff@jnl{Phys.Rev.B}}         
\def\prc{\reff@jnl{Phys.Rev.C}}         
\def\prd{\reff@jnl{Phys.Rev.D}}         
\def\prl{\reff@jnl{Phys.Rev.Lett}}      
\def\pasp{\reff@jnl{PASP}}              
\def\pasj{\reff@jnl{PASJ}}              
\def\qjras{\reff@jnl{QJRAS}}            
\def\skytel{\reff@jnl{S\&T}}            
\def\solphys{\reff@jnl{Solar~Phys.}}    
\def\sovast{\reff@jnl{Soviet~Ast.}}     
\def\ssr{\reff@jnl{Space~Sci.Rev.}}     
\def\zap{\reff@jnl{ZAp}}                        
\def\nat{\reff@jnl{Nature}}             

\title[Fullerenes in IC 348]{Fullerenes in the  IC 348 star cluster of the Perseus Molecular Cloud}
\author[S. Iglesias-Groth]{S. Iglesias-Groth$^{1,2}$\thanks{E-mail:
sigroth@iac.es}\\
$^{1}$ Instituto de Astrof\'\i sica de Canarias, La Laguna 38200, Spain\\
$^{2}$Departamento de Astrof\'\i sica de la Universidad de La Laguna, Avda. Francisco S\'anchez, La Laguna, 38200, Spain \\}
\begin{document}

\date{May 2019}

\pagerange{\pageref{firstpage}--\pageref{lastpage}} \pubyear{2018}

\maketitle

\label{firstpage}

\begin{abstract}

 We present the  detection of fullerenes C$_{60}$ and C$_{70}$ in the  star-forming  region IC 348 of the Perseus molecular cloud. Mid-IR vibrational transitions of  C$_{60}$ and  C$_{70}$ in emission are found  in  \textit {Spitzer} IRS spectra  of individual stars  (LRLL 1, 2, 58), in the  averaged spectrum  of three other cluster stars (LRLL 21, 31 and 67) and in spectra obtained at four interstellar locations distributed  across the IC 348 region. Fullerene bands  appear widely distributed in this  region  with higher strength in the lines-of-sight of  stars  at the core of the cluster.    Emission features consistent with three most intense bands of the C$_{60}^+$ and with  one of  C$_{60}^-$ are also found in several spectra, and if  ascribed  to these ionized species it would imply ionization fractions at 20 and 10 $\%$, respectively. The stars under consideration host protoplanetary disks, however  the spatial resolution of the spectra is not sufficient to disentangle the presence of fullerenes in them. If fullerene  abundances in the cloud were  representative of IC 348 protoplanetary disks, C$_{60}$, the  most abundant  of the two species,  could host  $\sim$ 0.1 $\%$  of the total available carbon in the disks.  This should  encourage dedicated searches in young disks with upcoming facilities as JWST. Fullerenes provide a  reservoir  of pentagonal and hexagonal carbon rings which could be important as building blocks of prebiotic  molecules. Accretion of these robust molecules in early phases of planet formation may contribute to the formation of complex organic molecules in young planets. 

\end{abstract}

\section{Introduction}
Since the discovery of fullerenes C$_{60}$ and C$_{70}$ by \citet{cami10} in  the young planetary nebula Tc1 using   spectra from \textit{ Spitzer Space Telescope},  mid-IR bands of C$_{60}$  have  been identified in various astrophysical environments  including reflection nebulae \citep{sellgren10},   planetary nebulae  \citep{G-H11}, protoplanetary nebulae \citep{zhang11}    and  post-AGB stars  \citep{gielen11}. The  C$_{60}^+$ cation  has also been detected  in many lines-of-sight of the diffuse interstellar medium  \citep{foing94, foing97, walker15, campbell16},  in one reflection nebulae \citep{berne13} and, in one protoplanetary nebulae \citep{I-G13}.   The  relative band intensities of C$_{60}$ in various sources appear better described by a thermal distribution than by a single-photon heating and fluorescent cooling model, but a successful explanation of band ratios and  ionization balance of fullerenes in the various environments is still lacking. For instance, contributions from chemically bonded fullerenes in the solid state resulting from co-deposition of C$_{60}$ and C on the surface of refractory grains or from the energetic processing of amorphous carbon may  have to be considered for a proper explanation of the observational data \citep{krasnokutski19}. C$_{70}$ bands are very rarely reported in the literature, thus there is far less insight  on its relative abundance to  C$_{60}$.  

Fullerenes are efficiently formed in vaporization experiments of graphite \citep{kratschmer90} and could also be formed in  asymptotic giant branch  stars (AGB)  where circumstellar molecular synthesis  is very rich. These  stable molecules are able to survive under the harsh conditions  of the ISM as suggested by observations in several young stellar objects  \citep{roberts12}  and by the ubiquitous presence of  the diffuse interstellar bands at  957.7 and 963.2 nm which are associated to  the cation C$_{60}^+$  \citep{walker15,campbell16}. A significant fraction of interstellar carbon could be located in the C$_{60}$ cation, although original estimates of order 0.9$\%$ have been recently revised to values down to 0.1 $\%$ by \citet{berne17} who used the new oscillator strengths measured by \citet{campbell16}. These cation bands were  also  detected in one protoplanetary nebula  by \citet{I-G13}  who  inferred   that  0.86$\%$  of the carbon abundance in the nebula was in the form of  the C$_{60}$ cation using the old oscillator strengths. Such  abundances are consistent with   fullerene concentration values  of order 0.05-0.1 ppm  which are similar to  some controversial claims in meteorites  \citep{becker97, becker94}, as debated by \citep{heymann97, hammond08}, and to the  values inferred from the UV bump at 217.5 nm under the assumption that this prominent feature of extinction is caused by fullerenes  \citep{I-G04}.  At such abundances, fullerenes can play an important role in interstellar chemistry \citep{omont16}.

The main driver for the formation of fullerenes is temperature see e.g. \citet{jager09}. The presence of  fullerenes in some planetary nebulae  where  bands of polycyclic  aromatic hydrocarbons (PAHs) are absent appears also consistent with laboratory studies showing that the  efficiency of fullerene formation is favoured in hydrogen-poor environments  \citep{cami10}. However, observations also show  the coexistence of fullerenes and PAHs  in  planetary nebulae  of the Magellanic clouds   and  other nebulae  \citep{sellgren10}.  It is therefore important to extend fullerene  studies to other hydrogen-rich  astrophysical contexts.  The material near very young stars   is particularly interesting in this respect, as numerous spectroscopic studies in the near, mid/far infrared and millimeter range  \citep[see][]{henning13}  reveal a  rich chemistry that may lead to the formation of a large variety of complex organic molecules in the disks of these stars.  Among  the various chemical ingredients detected are:  silicates, H$_2$, CO, ices and  organic molecules like PAHs which are common and remarkably abundant.  In the inner disk regions, Spitzer has detected a rich organic  chemistry with species in the warm gas (T=200-1000 K) including CO, CO$_2$, C$_2$H$_2$, CN and HCN  \citep{teske11}.  It is important to investigate  the presence of fullerenes  in  young star forming regions  and  reveal any potential contribution of these molecules to the  observed  organic chemistry.

 In this work, we report the  detection of C$_{60}$ and  C$_{70}$ in  material of the molecular cloud intervening in several lines-of-sight of the young  stellar cluster IC 348. This stellar system with an age $\sim$  2 Myr,  and distance to the Sun 316 pc \citep{herbig98},  is located at the eastern end of the    well known Perseus  molecular cloud complex \citep{cernis93,snow04,herbig98,luhman03}.  {\it Spitzer} MIPS maps of 24 and 70 $\mu$m  emission show that the IC 348 region contains a diffuse distribution of   warm dust most likely heated by soft-UV radiation \citep{bally08}.

\begin{table*}
\begin{minipage}{170mm}
\begin{center}
\caption{ Parameters of stellar targets }
\scriptsize{
\begin{tabular}{cccccccccccc}
\hline
Object  & RA & Dec & Sp.T  & T$_{eff} $ (K) & M$_{*}$(M$_{\odot}$)&
L$_{*}$(L$_{\odot}$)&R$_{*}$(R$_{\odot}$)& A$_{V}$ & Age(Myr)& AORKey & dist. to LRLL 1\\

LRLL 1  & 03:44:34.20   & +32:09:46.3 & B5V     & 15400$^h$       & 6.15$^l$    & 1660$^j$  &
& 3.1$^g$    & 0.4$^l$ & 25310464$^{a}$ & 0.04" \\
LRLL 2  & 03:44:35.36 & +32:10:04.6 & A2    &  8970 & 3.5$^i$  & 137$^f$   &5$^f$   & 3.2$^g$    & 3$^m$ & 22847744$^{n}$ &  23.5" \\
 & & & & & & & & &                                                             & 22968320$^{b}$ &  23.5" \\
 & & & & & & & & &                                                             & 22968064$^{b}$ &  23.4" \\
 & & & & & & & & &                                                             & 16269056$^{c}$ &  23.6" \\
 & & & & & & &  & &                                                            & 22967808$^{b}$ &  21.8" \\
 & & & & & & &  & &                                                            & 22961664$^{b}$ &  21.8" \\
LRLL 21 & 03:44:56.15 & +32:09:15.5 & K0   & 5240  &  1.8$^{e}$ &2.52-3.66$^g$ &           & 4.$^g$     &      & 22848768$^{n}$ &  280.4"\\  
LRLL 31 & 03:44:18.18 & +32:04:57.1 & G6   & 5700  &  1.5$^g$     & 4.3$^g$     & 2.1       & 8.8$^g$    &      & 22849280$^{n}$ & 353.9"\\
LRLL 58 & 03:44:38.56    & +32:08:00.7 & M1.3 & 3800   & 0.7$^{e}$ & 0.72$^g$ &2.1$^f$ &3.4$^g$ & 3$^m$&  22966016$^{b}$ &   119.2" \\
 & & & & & & &  &&                                                                                &16755456$^{d}$ &    120.6" \\
 & & & & & & &  &&                                                                                &22964224$^{b}$ &    119.4" \\
 & & & & & & &  &&                                                                                &22966272$^{b}$ &    119.2" \\
 & & & & & & &  &&                                                                                &22966528$^{b}$ &    119.6" \\
LRLL 67 & 03:43:44.62 & +32:08:17.9 & M0   & 3720  &  0.6$^i$     &0.48$^g$     & 1.6$^f$  & 2.4$^f$   &      & 22850816$^{n}$ &  635.8"\\
\hline
\end{tabular}
}

$^a$ Program ID 50043 (K. Misselt). $^b$ Program ID 40372 (J. Muzerolle). $^c$ Program ID 2  (J. R. Houck).
$^d$ Program ID 179 (N. Evans). $^e$ Siess et al.,2000. $^f$ Luhman et al.,2003. $^g$ Flaherty et al., 2012,$^h$ Merin et al. 2004, Kalas et al. 1997, $^i$  Siess et al. 2000,$^j$ Olofsson et al. 2012, $^k$ Preibisch et al. 2001,$^l$ Montesinos et al 2009, $^m$ Cohen et al.2004, $^n$ Program ID 40247 (N. Calvet), $^o$ Program ID 50560 (D. Watson).
\end{center}
\end{minipage}
\end{table*}

\begin{table*}
\begin{minipage}{170mm}
\begin{center}
\caption{ Location of IC 348 interstellar pointings}
\scriptsize{
\begin{tabular}{ccccc}
\hline
Interstellar Pointing  & AR & Dec &  AORKey & dist. from LRLL 1\\
p1  & 03:44:41.66 & +32:06:45.4 & 22848512$^{n}$ & 4"\\
p2 & 03:44:36.08 & +32:00:14.7 & 22850560$^{n}$ & 572" \\
p3 & 03:43:50.10 & +32:08:17.9 & 22851072$^{n}$ & 567" \\
p4 & 03:44:33.23 & +31:59:54.1 & 22851584$^{n}$ & 592" \\
\hline
\end{tabular}
}

$^n$ Program ID 40247 (N. Calvet).
\end{center}
\end{minipage}
\end{table*}

\section{Observations} 
\label{sec:data}

\subsection{Stars, circumstellar and interstellar  material}

 {\it Spitzer} IRS spectra  of several IC 348 stars obtained  in various programmes originally  aimed to  characterize the properties of protoplanetary  disks in stars of this young cluster and spectra of various interstellar locations distributed across the cluster have been examined with the goal to identify   fullerene emission bands. All the spectra under consideration in this work were taken at a distance less than 10 arcmin from the center of the  cluster (at physical distance of less than 1 pc in the projected sky). The main sample includes  the central  star  and most luminous member of IC 348 (LRLL 1,  \citealt{luhman98})  and  another two stars  with evolved disks LRLL  2 and LRLL 58   selected because of the availability of multiple observations  by \citet{flaherty12}. Their   parameters are listed in Table 1. Spitzer maps reveal that the IR emission surface brightness peak of the Perseus cloud coincides with the location of the B5 V star LRLL 1 (HD 281159), a binary star (separation 0.6 arcsec) with an extended disk structure mapped by \citet{olofsson12} using polarimetry.  The polarimetric measurements  reported by these authors and the ISO measurements at 16 $\mu$m show a disk structure extending more than 20 arcsec in the SE-NW direction (with less emission in the NW part). This structure was first identified by \citet{kalas97}  who proposed that it could be due to  a protoplanetary disk. This hypothesis was questioned by \citet{rebull07}, but  observations by  \citet{olofsson12} suggest that   the structure is indeed linked to circumstellar material near the star, which could be due to a protoplanetary disk or to the disruption of an accretion disk.

 The star LRLL 2 is an A2-type with  stable photospheric  flux and  near/mid IR excess   consistent with an optically thick dust at T$>$ 1000 K. LRLL 2 is a pre-main-sequence star with a bolometric luminosity of 137 L$_{\odot}$, a radius of 5 R$_{\odot}$   and a mass of  $\sim$3.5 M$_{\odot}$ based on the \citet{siess00}  3 Myr isochrones. \citet{espaillat12}   find that the  infrared SED can be fit with optically thick material  that extends inward to the dust destruction radius 1.7 AU.  \citet{flaherty12}  propose the existence of an evolved protoplanetary disk in LRLL 2 (see their Fig. 2).   Evolved disks are believed to lie between fully optically thick disks and debris disks.  Object LRLL 58 is a M1.3 star  \citep{siess00} which shows  H$\alpha$  emission clearly indicative of an ongoing accretion process.Details on its infrared variability gas and disk properties are provided by \citet{flaherty12}.      

 In addition to these three stars, we also discuss spectroscopic observations of several other   IC 348 stars (LRLL 21, 31 and 67) less intensively observed  than the previous two and of several interstellar locations distributed across the cluster. The individual spectra of these  three stellar targets  have modest S/N for detection of weak fullerene bands but we will show that averaging these spectra provides sufficient S/N to reveal the presence of many fullerene bands with high confidence. Similarly for the average of the four interstellar spectra. 																						
  
\subsection{Data} 

The observations of the IC 348 targets reported here were obtained with the InfraRed Spectrograph (IRS) \cite{houck04}  onboard the \textit {Spitzer Space Telescope}  \citep{werner04}. Targets  were selected by visual inspection of  the CASSIS atlas of spectra \citep{lebouteille11} which provides reduced  and flux calibrated background subtracted data (see details at http://cassis.sirtf.com/), particularly in  the spectral region 17-20 $\mu$m where the strongest C$_{60}$ vibrational bands are located.  The spectra  were obtained with the low resolution module-short wavelength (SL; 5-20 $\mu$m) and  with the high-resolution short wavelength module (SH; 9.5-19.5 $\mu$m) as part  of various independent programs. Details of the observational parameters, AORs and   the distance of the extracted slit position with respect to star LRLL 1  can be found in Tables 1 and 2. 

\begin{figure}
\includegraphics[angle=0,width=8cm,height=10cm]{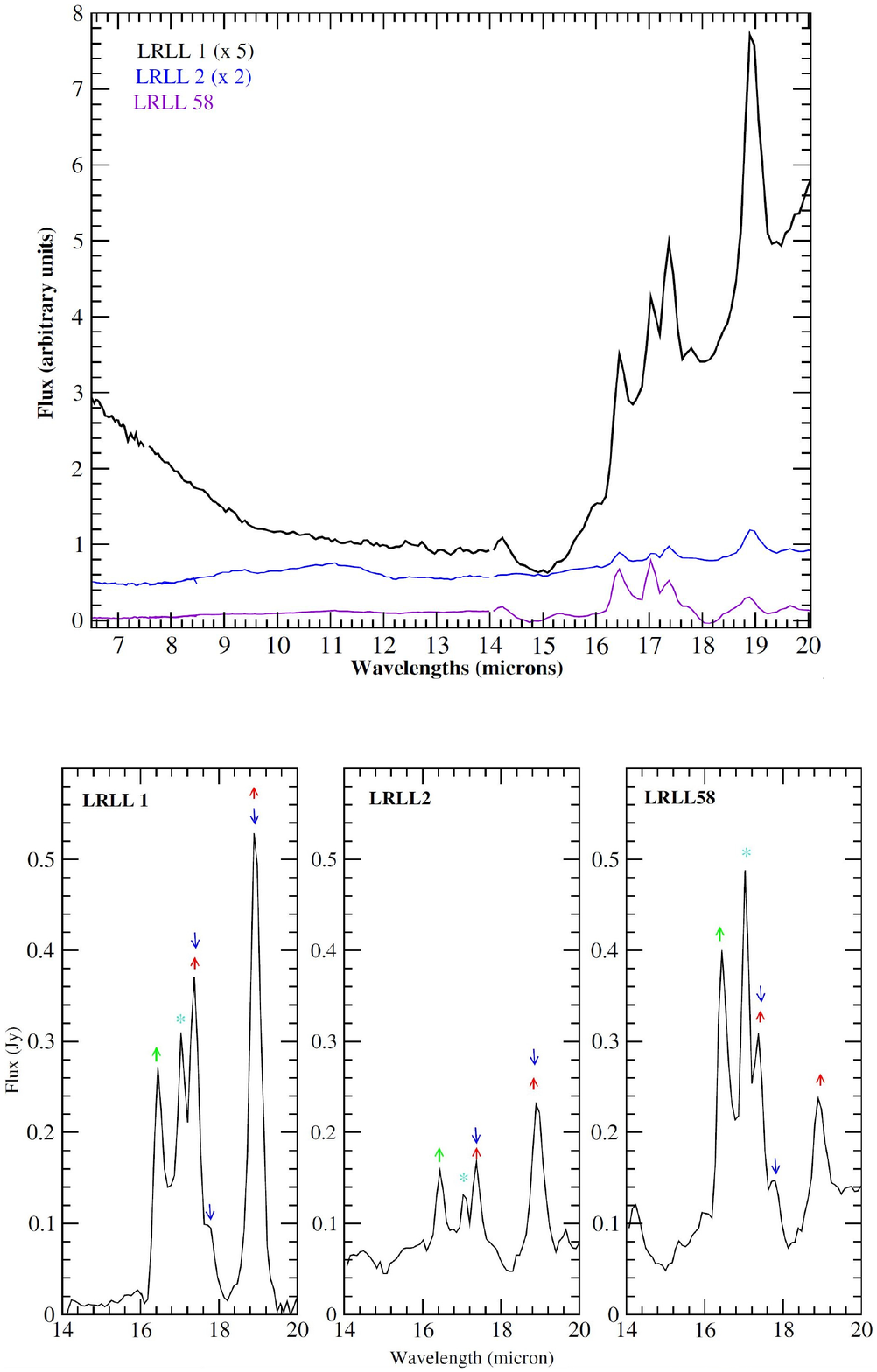}

\caption{Top panel: Mid-infrared low resolution spectra of the stars LRLL 1, LRLL 2 and LRLL 58 (AORs: 253104464, 16269056 and  16755456, respectively). Fluxes in Jy have been arbitrarily scaled for clarity in the display.  All spectra were obtained at distances less than 2"  from each of the stars. Bottom panel: Continuum subtracted spectra of  LRLL 1 (single spectrum), LRLL 2 (single spectrum) and LRLL 58 (average of 3 spectra obtained at distances less than 0.3 arcsec). Marks indicate the location of bands: C$_{60}$ (red),  C$_{70}$  (blue), H$_2$ (asterisk) and PAH (green) bands. The bands of C$_{60}$ and  C$_{70}$  are blended.
}
\end{figure}

The data processing is  described elsewhere (see http://cassis.sirtf.com/).  
The CASSIS extracted spectra  performs  the background subtraction using the  \textit {Adopt}  (Advanced Optimal Extraction) method described in detail by \cite{lebouteille10}. The particularity of this method is that the background profile is estimated for each row independently. \textit {Adopt} makes use of a polynomial background whose parameters are determined for each row solving a set of \textit {n} linear equations via  a multilinear regression algorithm. Two procedures are used (subtraction by order and by nod) to correct for the background, both provide similar results for our targets. The  column integrated spectra at the position of the stars  finally adopted here show the contribution of the emission lines in the intervening material toward the stars superposed to the stellar continuum. Spatially extended emission lines are seen at scales beyond the spatial profile of each of the  stellar targets.  The spatial scale of the IRS (approximately 5 arcsec per pixel) does not allow  to disentangle emission lines originating in the disk material from emission lines produced in the molecular cloud.  The spectrum extracted at the position of the star includes the contribution of the stellar photosphere, the  protoplanetary disk  and any  emission of the intervening cloud material. For the low resolution data the orientation of the SL and LL slits are different, but the slits overlap  on the target position and the extraction of the  spectrum at this position  minimize relative flux differences that may arise between the two observing modes due to a slightly different spatial location of the stellar profiles.

\begin{figure}
\includegraphics[angle=0,width=8cm,height=6cm]{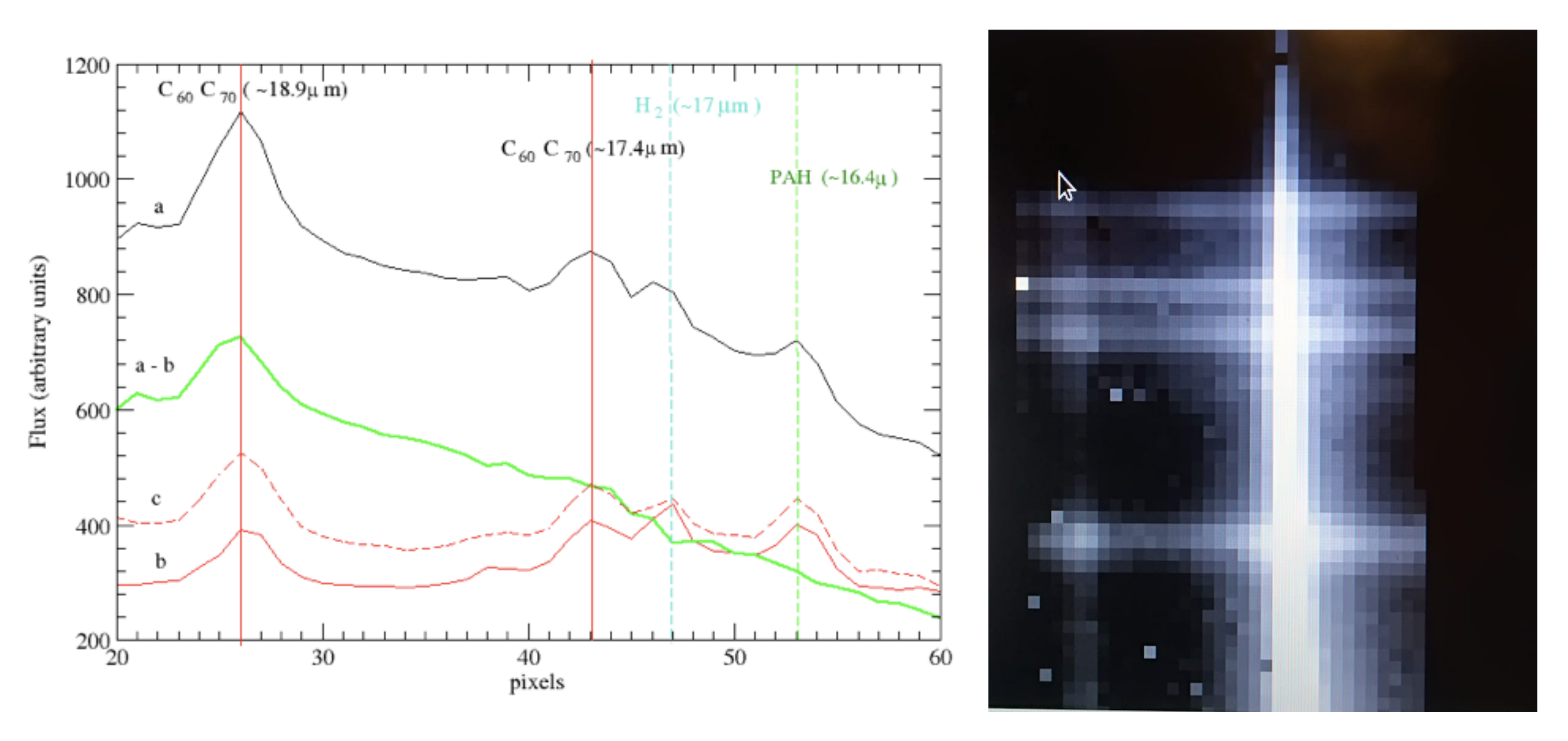}
\caption{  Right: Spectral 2-D image of LRLL1, the brightest star in the field. A second much fainter object is  located at the left part of the image. Several  emission lines are seen across the image. The strongest band at the bottom is the 18.9 $\mu$m C$_{60}$ band.  Left: Spectrum extracted at the position of  LRLL 1, as an average of the two columns with the highest recorded signal in the image  (label \textbf {a}, black line). For comparison, the  average spectrum extracted from  columns at the left of the star position (\textbf {b} red line), and to the right(\textbf {c} red dashed line) is plotted. The subtraction of spectra \textbf {a-b} is also plotted  (green line).  Marks and solid vertical lines color codes for bands:  C$_{60}$ (red), H$_{2}$ (blue)  and PAHs (green).
}
\end{figure}

\begin{figure}
\includegraphics[angle=0,width=8cm,height=12cm]{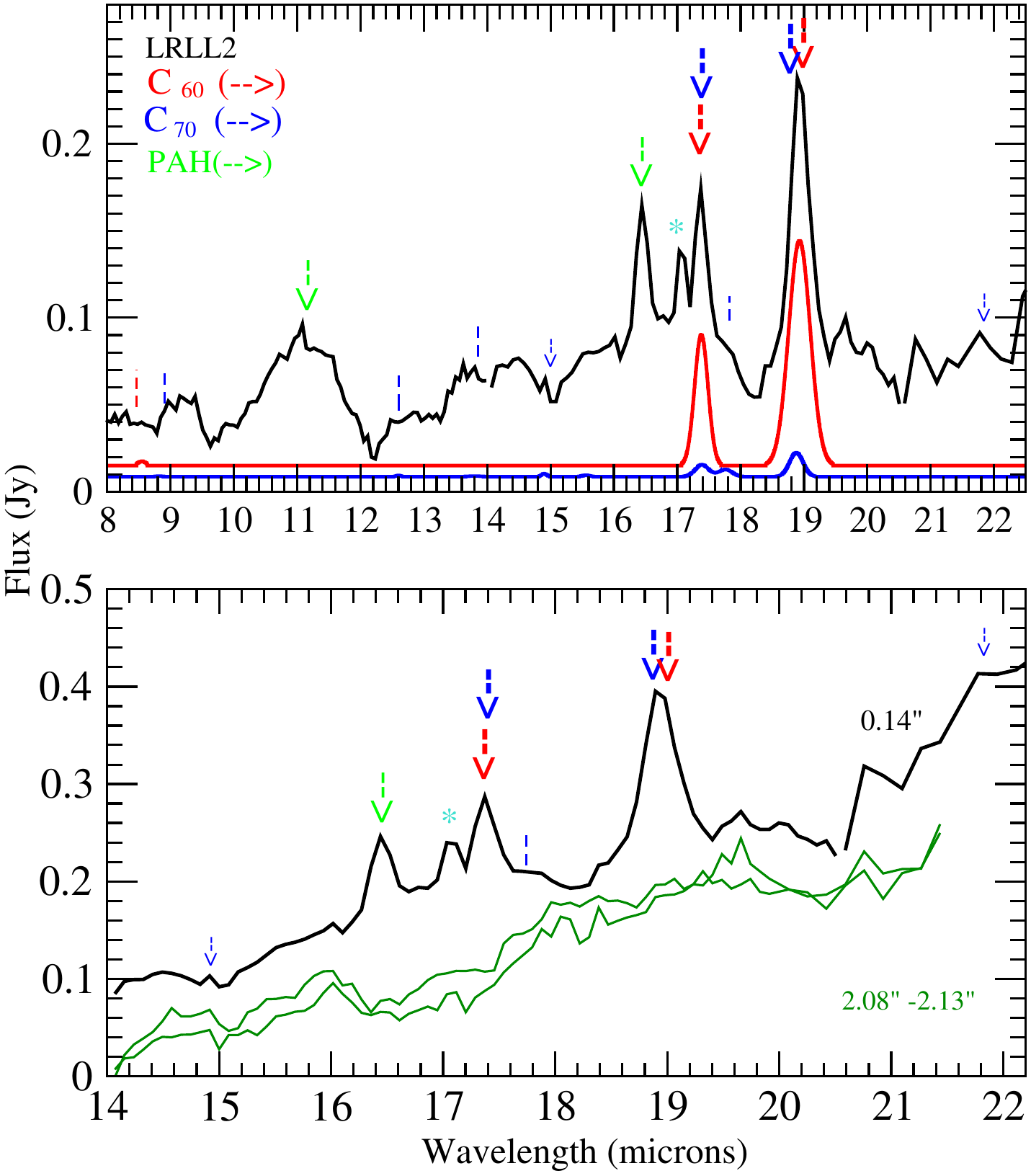}
\caption{Spectra of the A2-type star LRLL 2. Bottom panel:  spectrum obtained at a distance of  0.14"  from the star (black line) in comparison  with two spectra obtained at a distance of ~ 2" (green line).  Top panel:  simulation   of  C$_{60}$ and C$_{70}$ mid-IR bands in comparison with the observed spectrum. Marks and solid lines color codes:  C$_{60}$ (red), C$_{70}$ (blue)  and PAHs (green).
}
\end{figure}

\begin{figure*}
\includegraphics[angle=0,width=18cm,height=14cm]{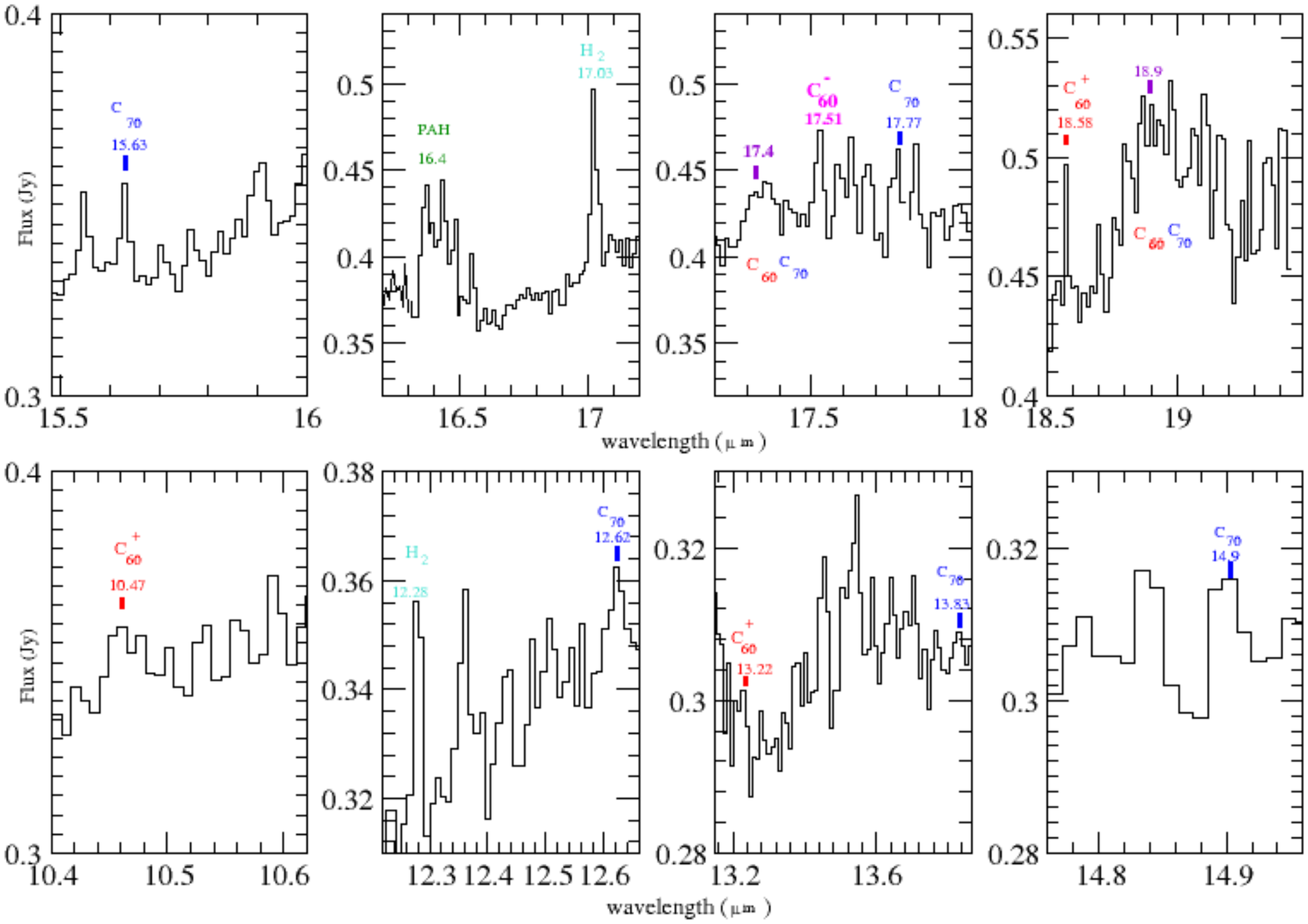}
\caption{Expanded view of the Spitzer mid-infrared high-resolution (HR) spectrum of the star LRLL 2 (AOR: 22847744)  }
\end{figure*}

\section{Results and discussion}

\subsection{Low resolution spectra}

Figure 1 (top panel) shows the low resolution Spitzer/IRS spectrum (7-20 $\mu$m) for three of the targets.  Note that in LRLL 1 the spectrum is dominated by a thermal dust continuum.   A most prominent emission band system is present in each star in the range 16-20 $\mu$m showing a remarkable band at 18.9 $\mu$m of C$_{60}$ (with potential contribution of C$_{70}$) and a mixture of PAHs at 16.4 $\mu$m, H$_2$ (17 $\mu$m) and fullerenes  C$_{60}$ and C$_{70}$ at 17.4 $\mu$m, see details in Fig. 1 (bottom panel)  where the bands are marked and the dust continuum was subtracted to LRLL 1.   The relative strength of the fullerene and PAH bands   changes notably from the hottest B5 V star (LRLL 1, bottom-left panel) and  the A2 V star (LRLL 2) where the fullerene bands are  the dominant  features, to the coolest star M1.3 (LRLL 58, bottom-right panel of Fig. 1),  where the PAH band is dominant. The spatial behaviour of  the emission bands in the intervening  material in the line of sight of each star varies notably. While the  bands are rather stable spatially in LRLL 2  and LRLL 58, in LRLL 1 inspection of the spatial evolution of the  C$_{60}$ band at 18.9 $\mu$m shows a stronger band closer to the spatial location of the star. In Fig. 2 we plot the LRLL 1 spectrum extracted from  various positions along the slit, the spectrum corresponding to the position of the star  is  labeled with ÒaÓ, we also plot average spectra extracted from nearby columns,  left and right from the star location,  labeled Òb,cÓ, respectively.  The difference spectrum Òa-bÓ shows that at the position of the star the strength of the 18.9 $\mu$m is significantly higher than in more distant regions. This could be interpreted as evidence for a higher excitation of  the C$_{60}$ molecules because of the higher UV radiation field near the star.

In Fig. 3  we present details of the bands and the continuum emission  of the A2-type star LRLL 2. At the bottom panel we compare spectra  where the bands of fullerenes and PAHs are clearly detected taken at the nominal position of the star (less than 0.14 arcsec)  with spectra taken at separation of  $\sim$ 2 arcsec where these bands are not detectable. The continuum emission remains similar at the two positions.  In the upper panel of Fig. 3, the spectrum has been subtracted for the continuum emission. The positions of known bands of  C$_{60}$ (8.5, 17.4  and 18.9 $\mu$m) and C$_{70}$ (17.4, 17.8, 18.9  and 21.8 $\mu$m) are marked. The solid red and blue lines in this panel represent computed simulations of the relative strengths of the bands of C$_{60}$ and C$_{70}$ which best fit the observed spectrum. The 8.5 $\mu$m band is too weak and blended with PAHs for a reliable detection. Weak features of C$_{70}$ could be present in the spectrum at 17.8 and 21.8 $\mu$m, but their  detection is   marginal and require confirmation with higher resolution measurements.  These bands  are  important  to ascertain the relative contribution of this molecule to the strong and broad bands at 17.4 and 18.9 $\mu$m.  

\subsection{High-resolution spectra}

In Fig. 4 we plot spectra of LRLL 2 obtained  with the  IRS  high spectral dispersion module (R=600).   The various panels of this figure display zooms of several spectral regions with the position of known individual bands of C$_{60}$, C$_{60}^+$, C$_{60}^-$ and C$_{70}$ marked and the observed wavelength of the peak emission for each band  indicated. In addition to the bands at 18.9 and 17.4 $\mu$m mentioned above, the high-resolution spectrum shows evidence for bands at 12.63,  13.83, 14.90, 15.63 and 17.77 $\mu$m which  can be attributed to C$_{70}$ vibrational  bands based on a comparison with  published  spectroscopic laboratory measurements.    \citet{I-G11}   measured laboratory wavelengths and FWHMs  for bands of C$_{60}$ and C$_{70}$  at different temperatures and  found evidence for wavelength shifts of these bands depending on  the  temperature  and  the adopted matrix. Table 3  lists for each band of C$_{70}$ the range of wavelengths measured at different laboratory conditions. The table also lists wavelengths for the  bands detected in  the IRS high dispersion spectrum  of LRLL 2 (typical measurement errors $\pm$ 0.01 $\mu$m).   Only bands in the range  10-20 $\mu$m covered by the  IRS spectrum are given in the Table. The agreement between laboratory wavelengths of C$_{70}$ and observed wavelengths for the five cleanest potential  bands of C$_{70}$  is very good and provides strong support for the identification of C$_{70}$ in LRLL 2. Other bands of C$_{70}$  are also present in the spectrum, for instance, those  measured at 17.4 $\mu$m and 18.7-18.9 $\mu$m but are  blended with bands of C$_{60}$ at very similar wavelengths. Complex broad features are detected in LRLL 2 at these wavelengths which could be contributed in principle  by  the two fullerene species.  The wavelengths of the  bands at 17.4 and 18.9 $\mu$m  are the most sensitive to temperature and matrix conditions according to  laboratory work. This   makes more difficult to disentangle the contribution of each fullerene specie  to these features. 

 The typical widths of  individual, clean C$_{70}$ bands measured in  the high dispersion spectrum of LRLL2 are 0.02-0.03 $\micron$ which is in  contrast wth the ten times broader features observed at 17.4 and 18.9 $\mu$m, however these features receive contributions from C$_{60}$ and C$_{70}$  and possibly also from bands of the ionized species, in particular C$_{60}^+$ is known to have  a relatively strong band at 18.6 $\mu$m, these features  are therefore intrinsically complex.  In addition, as noted in the laboratory work by \citet{I-G11}, the widths of C$_{60}$ bands are found  to be very sensitive to the temperature and the matrix conditions. In particular the 18.9  $\mu$m band which,  depending on the matrix and temperature conditions set at laboratory, displays a large range of widths spanning from 0.04-0.4 $\mu$m and also  wavelength shifts of up to 0.1 $\mu$m. The high dispersion spectrum of LRLL2  in the top right panel illustrates  the rather complex structure of the 18.6-19.0 $\mu$m feature, the weaker 17.4 $\mu$m feature is also complex.

In the high dispersion spectrum of LRLL 2 displayed in Fig. 4 there are  also features  at 10.47, 13.22 and 18.58 $\mu$m which are fully consistent with  wavelengths computed for the IR-active bands of C$_{60}^+$ by \citet{berne13} and therefore could be attributed to this cation. The bands at 10.47 and 18.58 $\mu$m are the strongest predicted for the spectral range covered by our observations  with intensities  2-3 times higher than the 13.22 $\mu$m band  (see  Table 2 in the previous paper) and in fact, this intensity ratio is consistent with the spectrum.  The other predicted lines are too weak and cannot be detected.  These three bands are the only ones  which could be detected in the high dispersion spectrum of LRLL 2. However in the low resolution spectrum it is possible to find features at 6.4 and 7.2 $\mu$m which according to predictions are the strongest of the IR active modes of the C$_{60}^+$.  In summary, the presence of  C$_{60}^+$ in the spectrum of LRLL 2 is  claimed based on identification of its five strongest IR active bands. 

In Fig. 4 we also show the detection of a clear band at 17.51 $\mu$m  which is consistent with one of the two strongest vibrionic bands  \citep{kupser08} of the anion C$_{60}^-$. In the low resolution spectrum of this star there is also a feature at 7.3 $\mu$m which could be due to the other strongest band of the anion. This second  feature is only marginally detected and therefore the presence of the anion should be taken with caution.

\begin{figure*}
\includegraphics[angle=0,width=18cm,height=14cm]{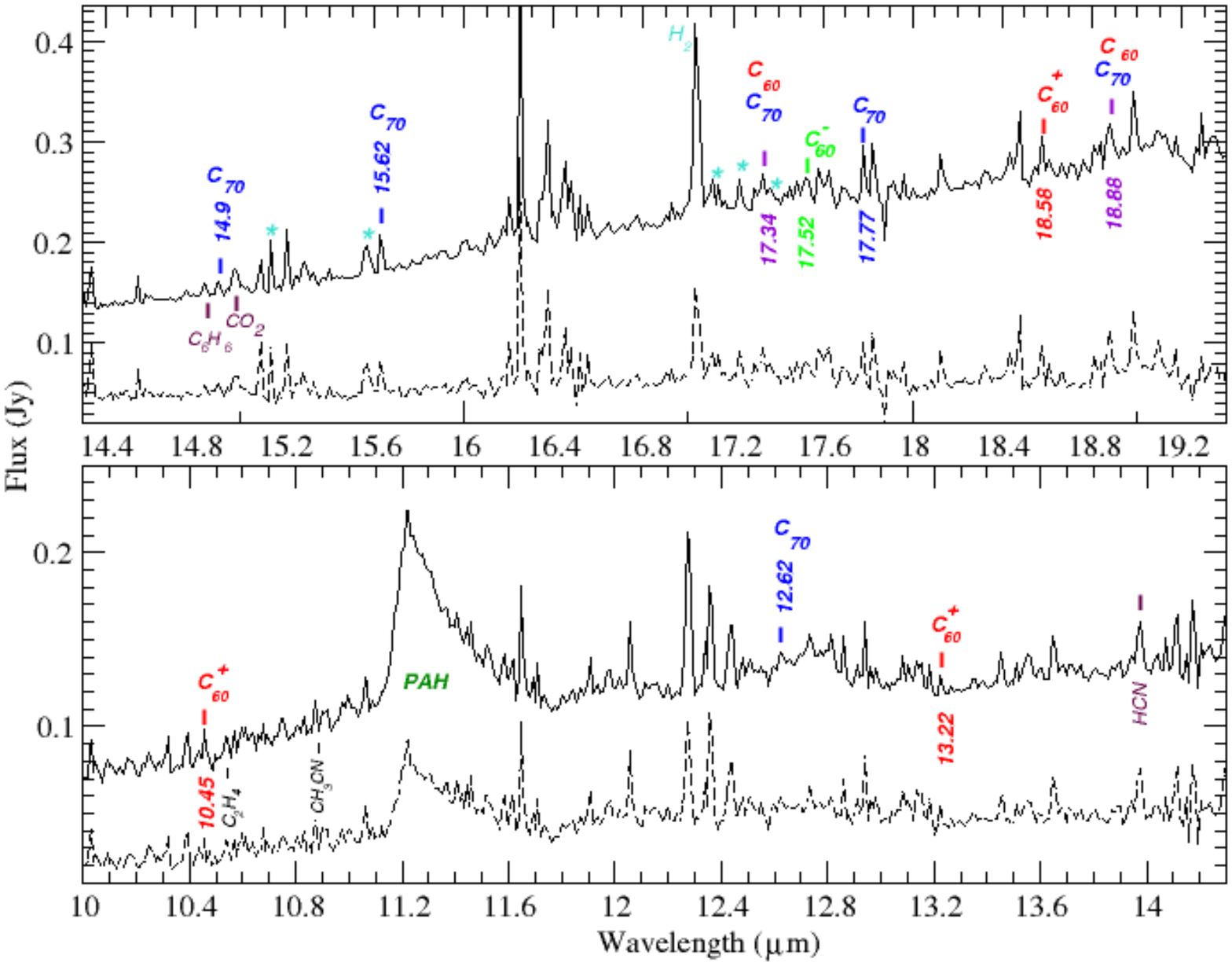}

\caption{Spitzer mid-IR averaged spectra  of three stars (LRLL 21, 31 and 67, solid line) and four interstellar
locations (dashed line) in IC348. The spectral range is 10-14.3 $\mu$m (bottom panel) and 10.4-19.3 $\mu$m (top panel). The location of bands of fullerenes and  some bands of  organic molecules   \citep{bast13} is indicated in both panels.  Some water bands \citep{blevins16} are marked with asterisks. Ticks in the vertical axis are spaced by 0.01 Jy.}

\end{figure*}

\begin{figure*}
\includegraphics[angle=0,width=18cm,height=14cm]{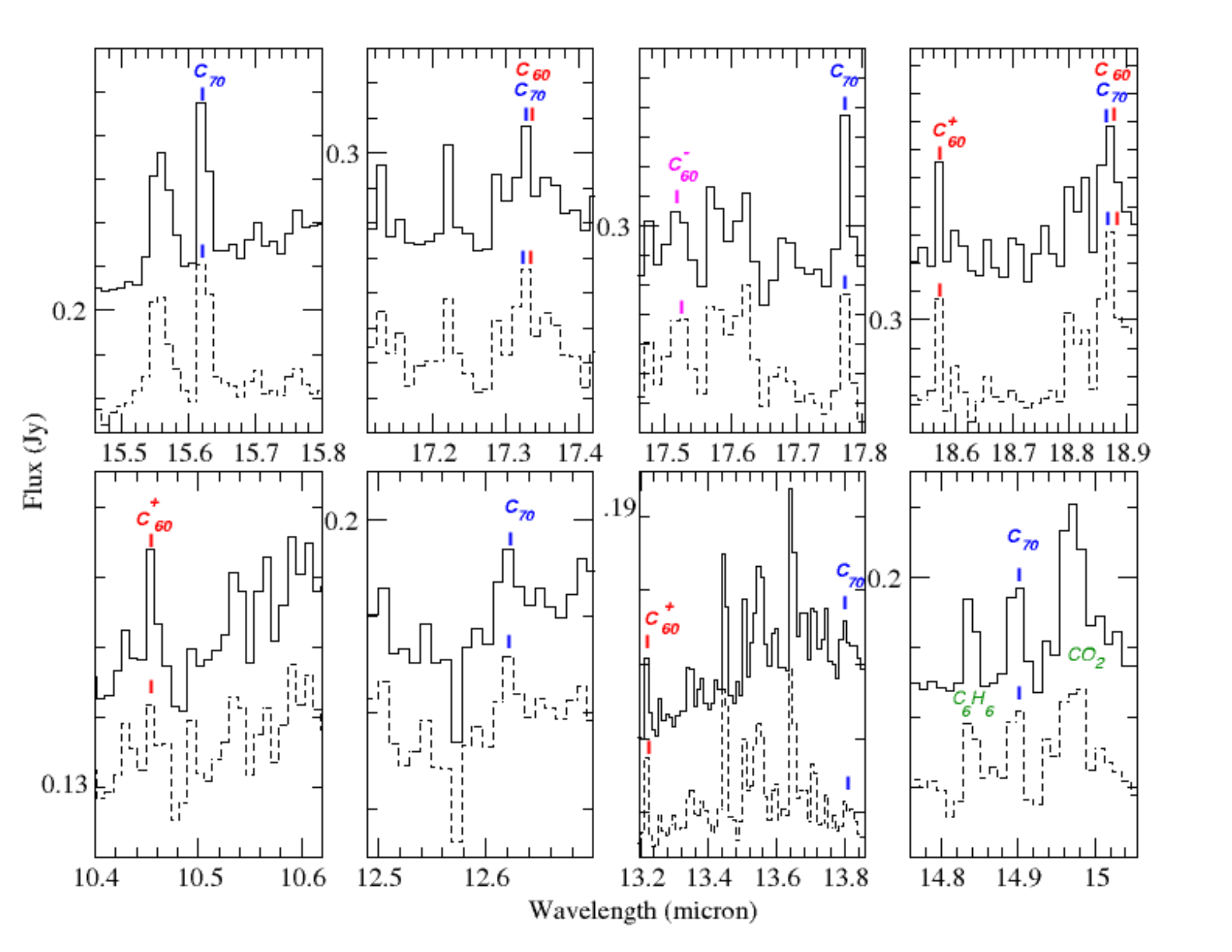}
\caption{Expanded view of Spitzer high-resolution  mid-IR averaged spectra of stars LRLL 21, 31 and 67 (top spectrum, continuous line) and four interstellar locations in IC348 (bottom spectrum, dashed line). }
\end{figure*}

 Other IRS high dispersion spectra of stars and interstellar locations
from IC 348  are displayed in Fig. 5 and 6 for comparison. The spectrum at the top
is the average of individual  spectra of three IC 348 stars and the spectrum at
the bottom  is the average of individual spectra acquired at  four interstellar
positions in the region of the cluster. Features like the PAH band at 11 $\mu$m and
the H$_2$ band at 17.4 $\mu$m band appear significantly different in the two
spectra.  The  fullerene  bands previously reported in LRLL 2 are also present
in these two  spectra. The bands of  C$_{60}$ at 17.4 and 18.9 $\mu$m;  
C$_{70}$ at 12.63,  13.83, 14.90, 15.63 and 17.77 $\mu$m;  C$_{60}^+$ at 10.47,
13.22 and 18.58 $\mu$m and C$_{60}^-$ at 17.51 $\mu$m are all clearly seen and
marked in the plots. Fig. 6 provides a zoom in the relevant spectral ranges to facilitate the identification and subsequent comparison. The strength of the fullerene bands appear very similar in both the average stellar and interstellar  spectra  indicating  that the emission bands are mainly  produced  in  molecular gas widely distributed in the interstellar medium of the cluster and therefore cannot be ascribed to the material in the disks of the selected stars.

\begin{table*}
\begin{minipage}{170mm}
\begin{center}
\caption{Laboratory wavelengths of C$_{70}$ bands  and observed bands in the high dispersion spectrum of LRLL 2}
\scriptsize{
\begin{tabular}{ccccc}
\hline
 &   Laboratory measurements  & LRLL 2 obs. spectrum & Background \\
 &($\mu$m)&($\mu$m) \\
\hline
&12.61-12.62					&	12.62  & 12.63\\
&13.80-13.81					&	13.83  & 13.81\\
&14.90-14.91					&	14.90  & 14.90\\
&15.64-15.67					&	15.63  & 15.63 \\
&17.62-18.07					&	17.77  & 17.78 \\
\hline
\end{tabular}
}

\end{center}
\end{minipage}
\end{table*}

\begin{figure}
\includegraphics[angle=0,width=9cm,height=7cm]{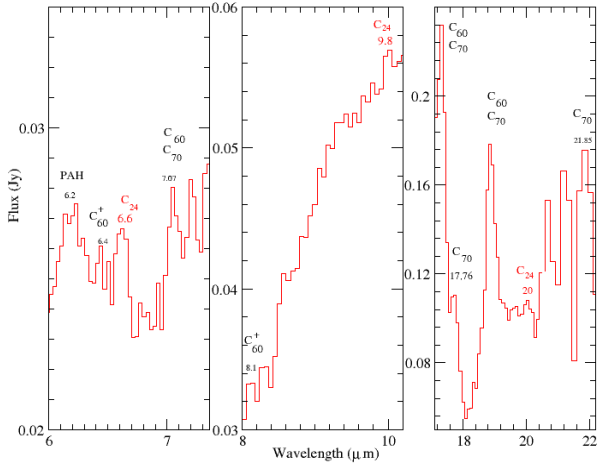}
\caption{Mid-IR spectrum of LRLL 58 Marks and color codes:fullerene  C$_{60}$ cation  (6.4, 
 8.1 $\mu$m),  C$_{70}$ (17.76,  21.85  $\mu$m), C$_{60}$+C$_{70}$ blends (7.1, 17.4, 18.9  $\mu$m), 
 graphene C$_{24}$ (red colour 6.6,9.8, 20 $\mu$m.
 }
\end{figure}

\begin{table*}
\begin{minipage}{170mm}
\begin{center}
\caption{ Measurements of fullerene bands in spectra of  IC 348 stars. Wavelength ($\mu$m) and Flux (W$m^{-2}$). From low-resolution  (LRLL1 and LRLL58) and high-resolution  (LRLL2) spectra. Flux statistical  errors are of order 20 $\%$.}
\scriptsize{
\begin{tabular}{cccccccccc}
\hline
Band &&  LRLL 1 &  & & LRLL 2 & & ~~  &LRLL 58 & \\
 &$\lambda$& Flux & FWHM & $\lambda$ &Flux & FWHM & $\lambda$ &Flux & FWHM \\
\hline
21.8 $\mu$m band & & &&&&&&&\\
~~C$_{70}$     & 21.78 &$\rm \leqq5x10^{-17}$  &           &       &                      &     & 21.85 & $\rm 4.5x10^{-17}$   &  \\
18.9$\mu$m band & & &&&&&&&\\
~~ \rm Total   & 18.93 & $\rm  1.1x10^{-15}$   & 0.33   & 18.93 &  $\rm 2.7 x10^{-16}$ & 0.3 & 18.87  &$\rm 4.0x10^{-16}$   & 0.35 \\
~~    C$_{60}$ &     & $\rm    9.8x10^{-16}$   &         &       &  $\rm 1.8x10^{-16}$   &     &       &$\rm 1.8x10^{-16}$     &  \\
~~    C$_{70}$ &     & $\rm    8.0x10^{-17}$   &         &       &  $\rm 7.3x10^{-17}$   &     &       &$\rm 1.5x10^{-16}$   &  \\

17.4 $\mu$m band &&&&&&&&& \\
~~ \rm Total   & 17.36 & $\rm 9.8x10^{-16}$    & 0.3     & 17.39 & $\rm 3.5x10^{-17}$   & 0.3 & 17.36 & $\rm 8.6x10^{-17}$   & 0.3 \\
~~ C$_{60}$    &       & $\rm 9.0x10^{-16}$      &        &       & $\rm 2.8x10^{-17}$   &     &       & $\rm 6.9x 10^{-17}$   & \\
~~ C$_{70}$    &       & $\rm 7.4x10^{-17}$       &        &       & $\rm 7.0x10^{-18}$   &     &       & $\rm  1.7x10^{-17}$ &  \\
17.76 $\mu$m band &&&&&&&&&\\
~~ C$_{70}$    & 17.76 &$\rm 9.6x10^{-17}$&$\leqq 0.3$  &  17.77 &$\rm 1.7x10^{-17}$ &$<$ 0.2    &  17.76 & $\rm 1.2x10^{-17}$ & $\leqq$0.3 \\
15.6 $\mu$m band & & &&&&&&&\\
~~C$_{70}$       &          &                   &     & 15.64        &$\rm 7.1x10^{-18}$ &      &  15.56   &  $\rm  2.8x10^{-17}$  &  \\
14.9 $\mu$m band & & &&&&&&&\\
~~C$_{70}$     & 14.95 &$\rm 7.5x10^{-18}$&        & 14.90  & $\rm 7.4x10^{-18}$& $\geq0.05$& 14.94 & $\rm 1.4x10^{-17}$ &  \\
12.6 $\mu$m band & & &&&&&&&\\
~~ C$_{70}$      &   &                   &     & 12.62        &$\rm 1.x10^{-17}$  &       &              &      \\
7.0 $\mu$m band & & &&&&&&&\\
~~ C$_{60}$+ C$_{70}$ & 7.1   &$\rm 1.5x10^{-17}$ &     & 7.04         &$\rm 1.9x10^{-17}$ &              & 7.07        & $\rm  2.0x10^{-17}$   & \\
\hline
\end{tabular}
}

\end{center}
\end{minipage}
\end{table*}

\begin{table*}
\begin{minipage}{170mm}
\begin{center}
\caption{ Measurements of fullerene bands  in the high-resolution averaged spectrum of LRLL 21, LRLL 31 and LRLL 67  and in the averaged interstellar spectrum of four locations in the region of the IC 348 cluster. Wavelength ($\mu$m) and Flux (W$m^{-2}$). Flux statistical errors are of order 20 $\%$.
}
\scriptsize{
\begin{tabular}{ccccccc}
\hline
Band &&   Average Discs &&& Background \\
& $\lambda$ &Flux & FWHM & $\lambda$ &Flux & FWHM \\
\hline
18.9$\mu$m band  &&& &&&\\
~\rm Total   & 18.88 & $\rm 8x10^{-17}$ & 0.3 & 18.88 & $\rm 9x10^{-17}$  & 0.3 \\
17.4 $\mu$m band &&& &&&\\
~ \rm Total  & 17.33 & $\rm 2x10^{-17}$ & 0.08 & 17.34 & $\rm 2x10^{-17}$ & 0.07 \\
17.76 $\mu$m band  &&& &&&\\
~ C$_{70}$   & 17.78 & $\rm 9x10^{-18}$ &0.014 & 17.78  & $\rm 5x10^{-18}$  & 0.011 \\
15.6 $\mu$m band &&& &&&\\
~C$_{70}$    & 15.63 & $\rm 8x10^{-18}$ &0.02 & 15.63  & $\rm 1x10^{-17}$ & 0.03 \\
 14.9 $\mu$m band &&& &&&\\
~C$_{70}$    & 14.90 & $\rm 4x10^{-18}$  &0.02 & 14.90 & $\rm 8x10^{-18}$ & 0.02  \\
12.6 $\mu$m band &&& &&&\\
~ C$_{70}$   & 12.63 & $\rm 5x10{-18}$ & 0.02 & 12.63 & $\rm 4x10^{-18}$ & 0.02 \\
\hline
\end{tabular}
}

\end{center}
\end{minipage}
\end{table*}

\begin{table*}
\begin{minipage}{170mm}
\begin{center}
\caption{ C$_{60}^+$ bands in LRLL 2 (LR and HR $>$10 $\mu$m), LRLL 58 (LR), stellar average (HR, $>$10 $\mu$m) and  interstellar average (HR $>$10 $\mu$m): measured wavelengths ($\mu$m) and fluxes (W m$^{-2}$).  Flux statistical errors are of order  20$\%$}
\scriptsize{
\begin{tabular}{cccccccc}
\hline
   LRLL 2     & &         LRLL 58       && Average (LRLL 21, 31 and 67)     &&  Interstellar avg. & \\
 $\lambda$ & Flux      & $\lambda$ & Flux              &$\lambda$ & Flux     & $\lambda$ & Flux \\

\hline
6.4     & $\rm  5x10^{-18}$ & 6.45 &  $\rm 5x10^{-18}$ &        &                      &        &                      \\
7.22    & $\rm \leqq 2x10^{-18}$ & 7.22 &  $\rm 1.4x10^{-17}$   &        &                      &        &                      \\
10.47   & $\rm  2.2x10^{-17}$ &      &                       & 10.46  & $\rm 5x10^{-18}$  &  10.46 & $\rm 2x10^{-18}$   \\
13.22   & $\rm  7x10^{-18}$ & 13.23 & $\rm 3x10^{-18}$   & 13.22  & $\rm 2x10^{-18}$ &  13.22 & $\rm 2x10^{-18}$  \\
18.56   & $\rm  1.4x10^{-17}$  &      &                       & 18.58  & $\rm 4x10^{-18}$  &  18.58 & $\rm 3x10^{-18}$  \\
\hline
\end{tabular}
}

\end{center}
\end{minipage}
\end{table*}

\subsection{Measurements. Abundance estimates}

 The S/N  of the high dispersion spectra of LRLL 2 and the two  averaged spectra  of stars and interstellar locations are similar. The detection of individual bands mentioned in the previous  subsection are at least with 5$\sigma$ confidence, many with a much higher confidence level as  the measurement of the  \textit{rms}  of the continuum  in various locations free of bands (13.3, 14.5, 15.6 and 16.7 $\mu$m) gives values around 0.003 Jy. 

 In Table 4 and 5 we provide wavelengths, fluxes in Wm$^{-2}$ and full width half maximum (FWHM) measurements  for the relevant  neutral fullerene C$_{60}$ and C$_{70}$  bands in the 7-22 $\mu$m regions. In Table 6 we provide  measurements for the bands (6.4, 7.2, 10.5, 13.2 and 18.6
$\mu$m) of the C$_{60}^+$. All the measurements have been performed using the SPLOT routine of  IRAF and its de-blending option was used  wherever it was required to disentangle multiple components, particularly in the region 16-18 $\mu$m. Gaussian profiles 
were assumed for all  the fits. Errors in the line fluxes were estimated  by propagating pixel error through the line integrals and are typically of order 20 $\%$. The minimum  flux we can detect in the high-resolution spectra corresponds to unresolved lines with a peak flux 3 times higher than the \textit{rms} and it corresponds to  $\sim$ 1 x 10$^{-18}$ Wm$^{-2}$. 

 The C$_{60}$ features in the range 17-19 $\mu$m have a typical width of 0.3-0.4 $\mu$m,  wider than the spectral resolution of the instrument and  similar to the widths observed in  planetary nebulae (like Tc 1). The total flux of the feature at 18.9 $\mu$m varies by a factor 10,  from values close to 1 x 10$^{-16}$ up to 1 x 10$^{-15}$  W m$^{-2}$. The contribution of C$_{70}$ to the total 
emission of the bands observed at 18.9 $\mu$m and  17.4 $\mu$m  has been established using the information provided by other bands of this molecule  assuming a simple model to describe 
band ratios,  we find C$_{70}$ contributions in the range  10-30$\%$ of the total band  strengths.  

As discussed e.g. by \citet{bernard-salas12}  the excitation of  fullerenes in interstellar gas may be understood  via thermal models or IR fluorescence models.   The latter predict nearly constant F(17.4 $\mu$m)/F(18.9 $\mu$m) ratios in a broad range of excitation energies and also  fairly strong bands at 7 and 8.5 $\mu$m.  Thermal models where the emitted power in each of the fullerene bands  is proportional to the corresponding excited vibrational state can naturally explain   observed flux ratios F(17.4 $\mu$m)/F(18.9 $\mu$m) in the range 0.5-0.6 and the $\sim$ 50 times weaker fluxes of the 7 and 8.5 $\mu$m bands w.r.t. the 18.9 $\mu$m band in our spectra  (see the diagnostic diagram of Fig.5 in  \citet{bernard-salas12}).

From the fluxes in Table 4, it is possible to  estimate  abundances of the fullerenes C$_{60}$ and  C$_{70}$ by using thermal models and  adopting  the temperature dependence of the absorptivity for each fullerene transition \citet{I-G11}.  This approach has been followed for instance by \citet{G-H11}  in their study of fullerene abundances in planetary nebulae. It is  assumed  errors of 10 $\%$ in the flux determinations of C$_{60}$ and of order 20 $\%$ for the weaker lines of C$_{70}$. In the case of LRLL 2, where both the low and high spectral resolution data are available, a good number of  C$_{70}$ bands are detected (12.6, 14.9 and 15.6 $\mu$m) which are essentially uncontaminated in the high dispersion spectrum. These bands are  the basis for an initial thermal model solution for this molecule. 
Then,  an iterative process is followed to obtain an estimation of the  C$_{70}$   fluxes at 18.9, 17.4 and 7.0 $\mu$m  from   a best-fit solution providing both  the temperature of the emitting region (T$\sim$ 200 K) and  the total number of emitting molecules, which is found at   N(C$_{70}$)=1.5 x 10$^{44}$.  The fit also provides  an estimate of the fluxes for each of the three blended bands. The values are  listed in Table 4 where  total fluxes measured  for these bands and the best estimate of the C$_{70}$ and C$_{60}$ contributions to the  total fluxes are given. As a result only 30 $\%$ of the flux at 18.9 $\mu$m is contributed by C$_{70}$.  The resulting  thermal model  for C$_{60}$ band fluxes gives N(C$_{60}$)= 3.2 x 10$^{44}$ molecules, a factor 2 more abundant than C$_{70}$ and equilibrium temperature of 250 K.   In addition, the five  measured bands of  C$_{60}^+$ (fluxes listed in Table 3) albeit weak and close to the sensitivity limit of the spectra, also provide  a reliable model solution,  resulting in   N (C$_{60}^+$) = 2.6 x 10$^{43}$ molecules. According to the thermal model,  this  implies that  about 10 $\%$ of the  C$_{60}$  would be in the cation form.

 The upper limits on the fluxes  of the 7 and 8.5 $\mu$m bands limit  our capacity to determine accurate equilibrium temperatures and abundances in the other two stars, but in the case of LRLL 58 where at least three bands  of each neutral fullerene type are also detected in low resolution spectra, the preferred model solutions lead to  N(C$_{60}$)= 1 x 10$^{45}$ molecules with a vibrational temperature of  T=150 K and N(C$_{70}$)= 1 x 10$^{45}$ molecules with vibrational temperature of 300 K. Very similar abundances of both species. It was also possible to measure some weak bands of  C$_{60}^+$ in this star (see Table 6) from which it is inferred  N(C$_{60}^+$)=2 x 10$^{42}$ molecules and T=350 K. Indicating a much lower ionization fraction  than in the previous star. 

 For LRLL 1  where only upper limits can be imposed to the flux of the 7 $\mu$m band, for C$_{60}$ it is found  equilibrium temperatures below 250 K  and abundance values of  4 x 10$^{46}$ molecules with uncertainties of order 50 $\%$.  This star, which is the most luminous of the three under study,  is by far the one with the strongest bands and the highest value of N(C$_{60}$) according to these models.

An alternative way to estimate fullerene abundances in our sources is via the assumption that all  the UV energy absorbed by fullerenes is released in the IR, which  would be correct if there were no other relaxation channels. This is valid for C$_{60}$ internal energies (see \citet{berne17}).  If we also assume  that the UV absorption cross section of the neutral and the cation are similar,  we  can obtain a quick estimate of the ionization fraction by adding the fluxes measured for IR bands of  these species in our targets. In the case 
of LRLL2  the total flux measured in  bands of C$_{60}^+$ is 5 x 10$^{-17}$ W  m$^{-2}$ and for neutral C$_{60}$ is 2.3 x 10$^{-16}$, therefore the ionization fraction of C$_{60}$ is  21$\%$,  higher than  the value given above from
thermal models. In Fig. 4 (see also Fig. 6) we also show the detection of a clear band at 17.51 $\mu$m with a flux of 1.8 x 10$^{-17}$ W m$^{-2}$  which we attribute to the anion C$_{60}^-$. In the low resolution spectrum of this star there is also a feature at 7.3 $\mu$m  with flux 3 x 10 $^{-17}$ W m$^{-2}$ which can also be due to the anion. These two bands are the strongest vibrionic bands known of the anion \citet{kupser08}. Taking the sum of the   fluxes emitted in these bands as an estimate of the total emitted IR flux by the cation, we infer an ionization fraction of  21 $\%$, which should be strictly taken as a lower limit.

Following e.g. \citet{berne17}, the total  IR intensity (W m$^{-2}$ sr$^{-1}$) emitted by C$_{60}$ molecules is  I$_{tot}$=n(C$_{60}$) x $\sigma_{UV}$ x G$_0$ x 1.2 x 10$^{-7}$ where n(C$_{60}$) is the column density of C$_{60}$,  $\sigma_{UV}$= 4.2 x
10$^{-16}$ cm$^2$, G$_0$ is the radiation field. For the Perseus molecular complex  is generally adopted   a value of G$_0$=1, see e.g. the low spatial resolution studies of AME  (Anomalous Microwave Emission) in Perseus \citet{genova-santos15}. However
as shown in the study by \citet{sun08} large variations in the far UV (FUV) radiation field can exist  in  the Perseus star-forming region  IC 348  where the stars of this work are located.  In Table 3 of the latter paper the variations of the intensity
of the FUV radiation field  are quantified using different approaches. Depending on the adopted method, the center of the cluster where star LRLL 1 (HD 281159) is located has   a factor 10-80  higher  intensity field  than the less intense regions of the star-forming region which are comparable to the general interstellar field.  I will adopt here an intermediate value of G$_0$= 45 for the radiation field in the vicinity of  LRLL 1 and G$_0$=20 for  stars LRLL 2 and 58. 

The total fluxes of C$_{60}$ IR bands obtained from  Table 4 are divided by the area subtended (in sr) by the spatial extension of each extracted spectrum in order to provide a  determination of I$_{tot}$.  The values  obtained for LRLL 1, LRLL2 and LRLL 58 are 4 x 10$^{-7}$, 2 x 10$^{-7}$ and 0.5 x 10$^{-7}$ W m$^{-2}$ sr$^{-1}$, respectively, giving C$_{60}$ column densities of 2, 1 and 0.5 x 10$^{14}$ cm$^{-2}$ for each target. Using the expression f$^{C_{60}}_C$= n(C$_{60}$) x 60 / (N(H) x [C])  where [C] is the carbon to hydrogen ratio  adopted  as 1.6 x 10$^{-4}$ by \citet{sofia04}, and taking  N(H)= 4.8 x 10 $^{22}$ from \citet{sun08} for the central part  of the IC 348 cluster, it yields abundances of the gas phase carbon locked in C$_{60}$ of 1.6 x 10 $^{-3}$,  0.8 x 10$^{-3}$ and  0.4 x 10$^{-3}$, for stars LRLL 1, 2 and 58, respectively.  These values of 0.16-0.04 $\%$ compare well with previously reported values in star-forming regions  \citep{castellanos14, berne17}. This value is of order of those  found in planetary nebulae  \citep{G-H11}   suggesting that a significant fraction of fullerenes formed in proto and  planetary nebulae can survive the harsh conditions of the interstellar medium and become  incorporated  into the  material of young star-forming regions like IC 348. 

The typical total  disk masses in IC 348 stars as measured from millimeter 
fluxes \citep{lee11} are in  the range 0.002-0.006 M$_{\odot}$. Assuming  for the disks of our stars a typical  mass of 0.004 M$_{\odot}$ and solar metallicity,  we can estimate the  total mass of  carbon in the disks to be of order 6-2 x 10$^{-7}$ M$_{\odot}$. Applying the derived percentage of carbon  locked in fullerenes, we estimate that the  disks in our stars contain between
 8  and   2 x 10$^{-9}$  M$_{\odot}$ in the form of C$_{60}$, corresponding to a number of C$_{60}$ molecules in the range 10$^{46}$ - 10$^{45}$.

\section{Conclusions} \label{sec:spect}

We present the detection of fullerenes in interstellar matter of the star-forming region IC 348 using spectroscopic data  obtained by the {\it Spitzer Space Telescope}.  Detection of  bands of C$_{60}$ at 7.04, 17.4 and 18.9 $\mu$m in low resolution spectra and bands of   C$_{70}$  at 12.6, 13.8, 14.9, 15.6, 17.8, 18.9 and 21.8 $\mu$m in high and low resolution spectra are reported.  The bands of PAHs (16.4 $\mu$m) and H$_2$ (17 $\mu$m)  coexist with fullerene bands in the spectra of the targets under study.  Contrary to PAHs, fullerenes appear to be little sensitive to the physical conditions of the intervening material.  We also detect several bands at wavelengths consistent with those of the ionized species C$_{60}^+$ (10.47, 13.22 and 18.58 $\mu$m) and C$_{60}^-$ (7.3 and 17.51 $\mu$m) from which we estimate ionization fractions of 20 and 10 $\%$, respectively for the best measured spectrum. Fullerenes present in the gas of the IC 348 star-forming region  could  be incorporated into  protoplanetary disks and ultimately into planets during  their early stages of formation. If the derived  percentage of gas-phase carbon locked in fullerenes  were also representative of the material conforming the protoplanetary disks in IC 348 stars, a large reservoir of fullerenes may exist in  disks and could be revealed by  observations with upcoming facilities such as JWST.   Fullerenes are robust molecules consisting of pentagonal and hexagonal carbon structures  that may  supply  building blocks for  the formation of complex organic molecules \citep{iglesias-groth13} in young planetary environments.

\section*{Acknowledgements}
Based on observations made with  the \textit {Spitzer Space Telescope}.  I  thank CASSIS for access to the spectroscopic database, the MINECO project ESP2015-69020-C2-1-R for financial support and Rafael  Rebolo for a critical reading of the manuscript, valuable  suggestions and comments on its contents.

\label{lastpage}
\end{document}